# Converging Shock Flows for a Mie-Grüneisen Equation of State

Scott D. Ramsey[1]
Emma M. Schmidt[1]
Zachary M. Boyd[2]
Jennifer F. Lilieholm[3]
Roy S. Baty[1]

[1]Applied Physics, Los Alamos National Laboratory, Los Alamos, NM 87545
[2]Department of Mathematics, University of California Los Angeles, Los Angeles, CA 90095
[3]Department of Physics, University of Washington, Seattle, WA 98195

## Abstract

Previous work has shown that the one-dimensional (1D) inviscid compressible flow (Euler) equations admit a wide variety of scale-invariant solutions (including the famous Noh, Sedov, and Guderley shock solutions) when the included equation of state (EOS) closure model assumes a certain scale-invariant form. However, this scale-invariant EOS class does not include even simple models used for shock compression of crystalline solids, including many broadly applicable representations of Mie-Grüneisen EOS. Intuitively, this incompatibility naturally arises from the presence of multiple dimensional scales in the Mie-Grüneisen EOS, which are otherwise absent from scale-invariant models that feature only dimensionless parameters (such as the adiabatic index in the ideal gas EOS). The current work extends previous efforts intended to rectify this inconsistency, by using a scale-invariant EOS model to approximate a Mie-Grüneisen EOS form. To this end, the adiabatic bulk modulus for the Mie-Grüneisen EOS is constructed, and its key features are used to motivate the selection of a scale-invariant approximation form. The remaining surrogate model parameters are selected through enforcement of the Rankine-Hugoniot jump conditions for an infinitely strong shock in a Mie-Grüneisen material. Finally, the approximate EOS is used in conjunction with the 1D inviscid Euler equations to calculate a semi-analytical, Guderley-like imploding shock solution in a metal sphere, and to determine if and when the solution may be valid for the underlying Mie-Grüneisen EOS.

## I. INTRODUCTION

Scale-invariant solutions of the inviscid compressible flow (Euler) equations have been thoroughly investigated since at least the 1940s (see, for example, Taylor[1], Sedov[2], Guderley[3], and Zel'dovich and Raizer[4]), and in the time since applied in a wide variety of contexts including inertial confinement fusion[5-8], double-detonation supernovae[9], and other high energy density physics applications. A common theme of many such efforts has been the additional assumption of an ideal gas equation of state (EOS) closure model. Given that this EOS contains only one dimensionless parameter (the adiabatic index or ratio of specific heats $\gamma$), it is in some sense tailor-made for the construction of scale-invariant solutions.

Calkin and Davis[10] appear to be among the first to recognize the limitations of the "adiabatic equation of state" in the context of scale-invariant imploding shock waves, noting,

> "The applications of the … analysis are, therefore, limited to those cases in which [the adiabatic equation of state] is at least a fair approximation. This includes in particular, of course, for $1 < \gamma \leq 5/3$, the case of a perfect gas; in addition it is hoped that with $\gamma \sim 3$, [the adiabatic equation of state] may be approximately true for various metals."

and

> "…results derived from the procedure we have adopted do at the least furnish qualitative insight into the problem, and must describe the limiting behavior correctly to the extent that the hydrodynamical idealization itself has validity."

More recently, authors such as Coggeshall[11-13] effectively unified families of the aforementioned scale-invariant solutions using symmetry analysis techniques (a notion also investigated using phase space analysis by Guderley[3], Sedov[2], Lazarus[14], and Meyer-ter-Vehn and Schalk[15]). From these developments has also arisen the related effort of determining the conditions under which the inviscid Euler equations may be expected to admit scale-invariant solutions. As demonstrated by Ovsiannikov[16], Holm[17], Axford and Holm[18], Hutchens[19], and most recently Ramsey et al.[20] and Boyd et al.[21-23], the existence of these solutions essentially amounts to the EOS realizing a particular form. The ideal gas thus turns out to be but one of a general class of EOS instantiations that admit scale-invariant solutions, when coupled to the inviscid Euler equations.

As may be expected on the grounds of physical intuition, scale-invariant EOS classes feature constitutive laws that are linear and homogeneous in the pressure variable. While this behavior may prove reasonable in describing wide variety of gaseous materials, it is generally inconsistent with even simple EOS models intended for use with solid materials. The stiffened gas and Mie-Grüneisen EOS models outlined by Harlow and Amsden[24] are two archetypal examples of this phenomenon, in that even in their analytical simplicity they fail to meet the outcome of scale-invariance noted above. Indeed, EOS models of this type admit only a severely limited but otherwise universal class of scale-invariant solutions (see, for example, Boyd et al.[21] and Ramsey et al.[20]), thus seemingly rendering irrelevant the possibility of their use in conjunction with the powerful outcomes of similarity or broader symmetry methods (as detailed principally by Barenblatt[25,26]).

There appear to be at least two natural paths toward reconciling the inconsistency between non-ideal materials models and the presence of similarity. The first of these is known as 'quasi-similar analysis', and essentially treats non-ideal behavior (e.g., in an EOS) as a perturbation about an idealized state. While variants of this method have been established and applied in related studies by Sakurai[27-29], Sedov[2], Hutchens[19], Ponchaut[30], and Sachdev[31], their use in the current context will be relegated to future work.

As an alternative to quasi-similar or related methods, the focus of this study will be the use of scale-invariant EOS forms to fit or otherwise approximate non-ideal forms that are not *a priori* amenable to the presence of scale-invariant solutions to the inviscid Euler equations (e.g., the Mie-Grüneisen EOS). With an approximate, scale-invariant EOS available, a wide variety of scale-invariant solutions to the inviscid Euler equations become immediately available, and can ostensibly be used to infer a variety of possible or limiting wave motions in archetypal solids. It is expected that these outcomes may prove useful to communities engaged in research pertaining to impact phenomena and other high strain rate flows.

In support of these goals, a brief review of the underlying mathematical model is provided in Sec. II (including a discussion of the adiabatic bulk modulus corresponding to a Mie-Grüneisen EOS, through which the scale-invariant approximations will be constructed). The necessary EOS condition for the existence of scale-invariant solutions of the inviscid Euler equations will be reviewed in Sec. III, and an example approximate EOS constructed. The approximate EOS will be used to calculate a Guderley-like converging shock solution in Sec. IV. Conclusions and recommendations for future work are provided in Sec. V.

## II. GAS DYNAMICS

Following Harlow and Amsden[24], the partial differential equations governing the one-dimensional (1D) symmetric motion of an inviscid, compressible fluid are

$$\frac{\partial \rho}{\partial t} + u\frac{\partial \rho}{\partial r} + \rho\left(\frac{\partial u}{\partial r} + \frac{ku}{r}\right) = 0\,, \tag{1}$$

$$\frac{\partial u}{\partial t} + u\frac{\partial u}{\partial r} + \frac{1}{\rho}\frac{\partial P}{\partial r} = 0\,, \tag{2}$$

$$\frac{\partial e}{\partial t} + u\frac{\partial e}{\partial r} + \frac{P}{\rho}\left(\frac{\partial u}{\partial r} + \frac{ku}{r}\right) = 0\,, \tag{3}$$

where the mass density $\rho$, radial flow velocity $u$, pressure $P$, and specific internal energy (SIE; internal energy per unit mass) $e$ are regarded as functions of the radial position $r$ and time $t$, and $k$ = 0, 1, or 2 for planar, cylindrical, and spherical symmetry, respectively.

Using the fundamental thermodynamic relation[32-35] (a combination of the first and second laws of thermodynamics) between $\rho$, $P$, $e$, the fluid temperature $T$, and the fluid entropy $S$,

$$de = T\,dS + \frac{P}{\rho^2}\,d\rho\,, \tag{4}$$

and the chain rule, Eq. (3) becomes

$$\frac{\partial S}{\partial t} + u\frac{\partial S}{\partial r} = 0\,, \tag{5}$$

the equation for isentropic flow; this result is expected as dissipative processes (e.g., viscosity, heat conduction) are absent from Eqs. (1)-(3). Moreover, if the fluid entropy $S$ is assumed to be a function of the fluid density $\rho$ and pressure $P$, again using the chain rule Eq. (5) may be expanded to yield

$$\frac{\partial S}{\partial \rho}\left(\frac{\partial \rho}{\partial t} + u\frac{\partial \rho}{\partial r}\right) + \frac{\partial S}{\partial P}\left(\frac{\partial P}{\partial t} + u\frac{\partial P}{\partial r}\right) = 0\,, \tag{6}$$

or, with Eq. (1),

$$\frac{\partial P}{\partial t} + u\frac{\partial P}{\partial r} + K_S\left(\frac{\partial u}{\partial r} + \frac{ku}{r}\right) = 0\,, \tag{7}$$

where the function $K_S(\rho,P)$ is known as the inverse compressibility or adiabatic bulk modulus of the fluid. If $K_S$ is known, Eqs. (1), (2), and (7) are a closed system of three partial differential equations in the three unknowns $\rho$, $u$, $P$.

### II.A. Adiabatic Bulk Modulus

With Eqs. (6) and (7), the adiabatic bulk modulus $K_S$ is seen to be defined by

$$K_S\left(\rho,P\right) \equiv -\rho\,\frac{\left.\dfrac{\partial S}{\partial \rho}\right]_P}{\left.\dfrac{\partial S}{\partial P}\right]_\rho}\,, \tag{8}$$

and is also immediately recognizable as the inverse (adiabatic) compressibility as described by Callen[36] and Menikoff and Plohr[37]. It is also related to the fluid (adiabatic) sound speed $c$ by

$$K_S = \rho c^2 . \tag{9}$$

Moreover, $K_S$ may also be computed from an equation of state (EOS) given in the form $P = P(\rho,e)$. The chain rule may be used to expand the quantity

$$\frac{\partial P}{\partial t} + u\frac{\partial P}{\partial r} = \left(\frac{\partial \rho}{\partial t} + u\frac{\partial \rho}{\partial r}\right)\frac{\partial P}{\partial \rho} + \left(\frac{\partial e}{\partial t} + u\frac{\partial e}{\partial r}\right)\frac{\partial P}{\partial e}, \tag{10}$$

or, with Eqs. (1) and (3),

$$\frac{\partial P}{\partial t} + u\frac{\partial P}{\partial r} + \left(\rho\frac{\partial P}{\partial \rho} + \frac{P}{\rho}\frac{\partial P}{\partial e}\right)\left(\frac{\partial u}{\partial r} + \frac{ku}{r}\right) = 0, \tag{11}$$

from which, with Eq. (7),

$$K_S = \rho\frac{\partial P}{\partial \rho} + \frac{P}{\rho}\frac{\partial P}{\partial e}. \tag{12}$$

Equation (12) may thus be used to determine the adiabatic bulk modulus corresponding to an EOS of the form $P = P(\rho,e)$; if $K_S$ is instead specified, the corresponding EOS may also be determined from Eq. (12) using the Method of Characteristics.

## II.B. Mie-Grüneisen Equation of State

Variants of what is now called the Mie-Grüneisen (M-G) EOS were first constructed by G. Mie[38] and E. Grüneisen[39] in the early 1900s, and in the time since it has become archetypal as an analytical constitutive law relevant to the shock compression of a wide variety of crystalline solids. Thorough discussions of this EOS, its properties, and its use with compressible flow codes are provided among many others by Harlow and Amsden[24], Meyers[40], Gathers[41], Menikoff[42], and Segletes[43].

A broadly applicable form of the M-G EOS can be derived by considering shock propagation in the context of Eqs. (1)-(3). In the presence of a shock wave (regarded as a mathematical discontinuity for inviscid flows), the conservation of mass, momentum, and total energy across it is guaranteed by the application of the Rankine-Hugoniot 'jump conditions' at the position of the shock wave. These relations may be derived from the conservation form of Eqs. (1)-(3) (see, for example, Zel'dovich and Raizer[4]), and in a stationary-observer reference frame are given by

$$-u_{\mathrm{s}}\rho_1 = \left(u_{\mathrm{p},2} - u_{\mathrm{s}}\right)\rho_2 , \tag{13}$$

$$P_2 - \rho_1 u_{\mathrm{s}} u_{\mathrm{p},2} = 0 , \tag{14}$$

$$\frac{1}{2}u_{\mathrm{s}}^2 = e_2 + \frac{P_2}{\rho_2} + \frac{1}{2}\left(u_{\mathrm{p},2} - u_{\mathrm{s}}\right)^2, \tag{15}$$

where the subscripts 1 and 2 denote the material state immediately ahead of ('unshocked') and behind ('shocked') the shock wave, respectively, and $u_{\mathrm{p}}$ and $u_{\mathrm{s}}$ denote the particle (i.e., bulk flow) and shock velocities, respectively. In Eqs. (13)-(15), the common assumptions $P_1 = e_1 = u_{\mathrm{p},1} = 0$ have been assumed.

A large body of empirical data (see, for example, Meyers[40] or Cooper[44]) shows that the shocked particle and shock propagation velocities are approximately linearly related for a wide variety of crystalline solids,

$$u_{\mathrm{s}} = c_{\mathrm{ref}} + s u_{\mathrm{p},2}, \tag{16}$$

where $c_{\mathrm{ref}} > 0$and $s > 0$ (but typically in the range 1.0-1.7) are empirical fitting constants specific to a material. With Eq. (16), Eqs. (13)-(15) may be solved for $P_2$, $u_{\mathrm{p},2}$, and $e_2$ as functions of $\rho_2$,

$$P_{\mathrm{H}} = \frac{\rho_{\mathrm{ref}} c_{\mathrm{ref}}^2 \eta(\eta - 1)}{\left[\eta - s(\eta - 1)\right]^2}, \tag{17}$$

$$u_{\mathrm{p,H}} = \frac{c_{\mathrm{ref}}(\eta - 1)}{\eta + s(1 - \eta)}, \tag{18}$$

$$\begin{aligned} e_{\mathrm{H}} &= \frac{c_{\mathrm{ref}}^2 (\eta - 1)^2}{2\left[\eta - s(\eta - 1)\right]^2} \\ &= \frac{P_{\mathrm{H}}(\eta - 1)}{2\rho_{\mathrm{ref}}\eta}, \end{aligned} \tag{19}$$

where $\rho_1$ has been relabeled as $\rho_{\mathrm{ref}}$, $\eta = \rho/\rho_{\mathrm{ref}}$, and the '2' subscripts have been dropped for notational brevity. Equations (17)-(19) are referred to as the principal $P$-$\rho$, $u_{\mathrm{p}}$-$\rho$, and $e$-$\rho$ Hugoniots, respectively. Given Eqs. (17) and (19), perturbative deviations about the Hugoniot state may be constructed using the Taylor expansion

$$P = P_{\mathrm{H}} + \left.\frac{\partial P}{\partial e}\right]_{\rho} (e - e_{\mathrm{H}}) + \dots, \tag{20}$$

where

$$\left.\frac{\partial P}{\partial e}\right]_{\rho} \equiv \rho\Gamma(\rho), \tag{21}$$

and $\Gamma$ is referred to as the Grüneisen parameter. When Eq. (20) is truncated at first order, it becomes a first order M-G EOS:

$$P = P_{\mathrm{H}} + \rho\Gamma\left(e - e_{\mathrm{H}}\right). \tag{22}$$

As indicated by Eq. (21), the Grüneisen parameter is in general regarded as a function of $\rho$. The exact specification of this dependence may be motivated by theoretical or empirical considerations; various formulations and attendant commentary is provided by Harlow and Amsden[24], Menikoff[42], Axford and Holm[18], Axford[45], and Segletes[43]. Of particular interest to this work is the Segletes modification of the Dugdale-MacDonald parameterization,

$$\Gamma = \frac{2\left(2s-1\right)}{1+\eta}, \tag{23}$$

which is again appropriate for a wide variety of crystalline solids, and has been modified to ensure the existence of stable single shocks up the maximum density ratio allowed by Eq. (22).

With Eqs. (12), (17), (19), (22), and (23), the adiabatic bulk modulus corresponding to the M-G EOS is thus

$$K_S = \frac{\left(4s-1\right)P}{1+\eta} + \frac{2\rho_{\mathrm{ref}} c_{\mathrm{ref}}^2 \eta^2}{\left(1+\eta\right)\left[\left(1-s\right)\eta + s\right]^2}. \tag{24}$$

and is depicted in Fig. 1.

Equation (24) has a variety of notable features. It is unbounded for the density ratio given by

$$\eta_{\max} = \frac{s}{s-1}, \tag{25}$$

for any $P$; Eq. (25) indicates that $\eta_{\max}$ is the maximum achievable density ratio (including through isentropic processes). The minimum density ratio is $\eta = 1$ (where $\eta < 1$ corresponds to a rarefaction shock, which will not be considered in this work), where

$$K_S\left(\eta = 1\right) = \frac{\left(4s-1\right)P}{2} + \rho_{\mathrm{ref}} c_{\mathrm{ref}}^2, \tag{26}$$

$$\frac{\partial K_S}{\partial \eta}\left(\eta = 1\right) = \frac{4s-1}{4}\left(-P + 2\rho_{\mathrm{ref}} c_{\mathrm{ref}}^2\right). \tag{27}$$

Finally, Eq. (24) has a minimum in $\eta$ for any $0 \le P < \infty$. While the $\eta$-position of this minimum as a function of $P$ will not be reproduced here due to its being algebraically cumbersome, it is observed to occur at $\eta = 1$ for $P = 0$, and at $\eta \to \eta_{\max}$ as $P \to \infty$.

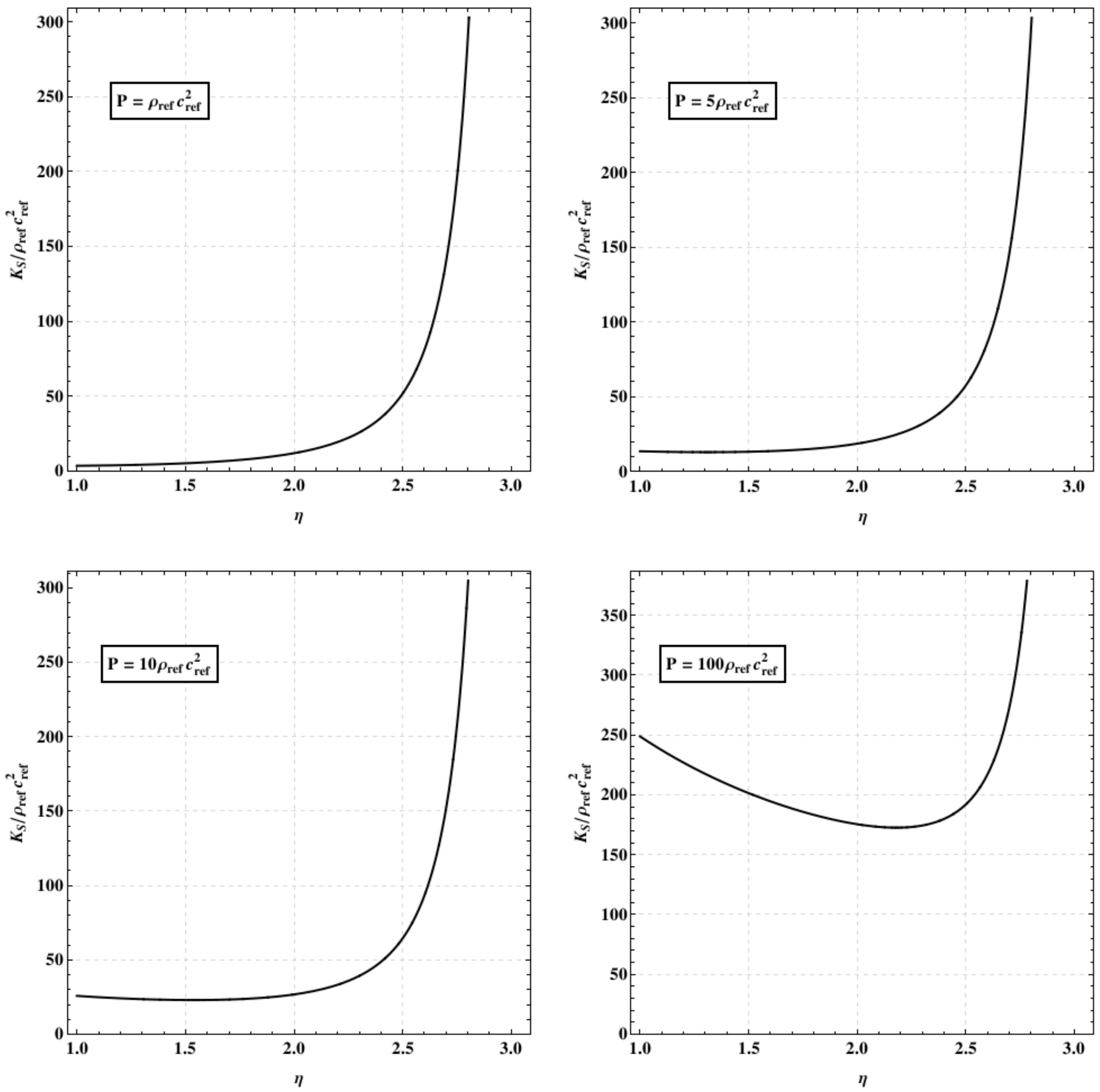

**Figure 1: Adiabatic bulk modulus for M-G EOS (Segletes modification of Dugdale-MacDonald parameterization), *s* = 1.489.**

## III. APPROXIMATE EQUATION OF STATE

Equations (1), (2), and (7) possess a wide variety of symmetry solutions when used in conjunction with the ideal gas EOS (see, for example, Coggeshall[11-13]),

$$P = (\gamma - 1)\rho e, \tag{28}$$

corresponding to, with Eq. (12),

$$K_S = \gamma P, \tag{29}$$

where $\gamma$ is interpreted as the dimensionless adiabatic index or specific heat ratio. Following Holm[17], Axford and Holm[18], Hutchens[19], Axford[45], Ovsiannikov[16], and Boyd et al.[21], symmetry analysis techniques may be used to show that this form of $K_S$ may be generalized to

$$K_S = P\varphi(\eta), \tag{30}$$

for the continued existence of scale-invariant solutions. In Eq. (30), $\varphi$ is an arbitrary function of the indicated argument. Equation (12) may then be used to recover the corresponding scale-invariant EOS,

$$P = (e - e_{\mathrm{ref}})\rho_{\mathrm{ref}}\eta\Gamma(\eta), \tag{31}$$

where $e_{\mathrm{ref}}$ is an arbitrary integration constant, and the Grüneisen parameter $\Gamma$ is regarded as a dimensionless function that must satisfy

$$\varphi(\eta) = 1 + \Gamma(\eta) + \eta\frac{\Gamma'(\eta)}{\Gamma(\eta)}. \tag{32}$$

where the prime denotes differentiation with respect to $\eta$. As noted in Sec. I, Eq. (31) is separable and homogeneous in $P$ and $e$ (up to the translation in $e$). It also includes one inherent dimensional scale (the parameter $\rho_{\mathrm{ref}}$).

Examples of scale-invariant solutions corresponding to Eq. (30) include the classical Noh problem[46] (which features a constant unshocked velocity field), and the classical Sedov[1,2] and Guderley[3,14] problems (both of which feature constant unshocked density fields). Examples of scale-invariant solutions wherein Eqs. (30) and (31) may assume more general forms include variants of the Sedov and Guderley problems featuring spatially-variable unshocked density fields, and the collapsing cavity problem (see, for example, Korobeinikov[47], Kamm[48], or Lazarus[14]).

In scenarios featuring the inclusion of multiple dimensional scale parameters (e.g., a density scale $\rho_{\mathrm{ref}}$ and velocity scale $c_{\mathrm{ref}}$), Eqs. (30) and (31) collapse to trivial forms, and no scale-invariant solutions of Eqs. (1), (2), and (7) can be constructed except in highly-specialized cases featuring planar symmetry ($k = 0$). These universal symmetry solutions are discussed at length by Boyd et al.[21-23] and Ramsey et al.[20], and will not be considered further in this work. Emphasis will instead be placed on Eqs. (30) and (31), and scale-invariant shock solutions existing within that context.

### III.A. Adiabatic Bulk Modulus

Comparison of Eqs. (24) and (30) immediately reveals that the M-G EOS as formulated in Sec. II.B is not of the scale-invariant form; thus when coupled to Eqs. (1), (2), and (7) via the adiabatic bulk modulus it will not prove compatible with the existence of scale-invariant solutions. The purpose of this work is to reconcile this inconsistency by approximating Eq. (24) using a judiciously selected form of Eq. (30).

Given that the pressure dependence of Eq. (30) is already specified, only the form of the arbitrary function $\varphi$ is available for selection in order to approximate Eq. (24). As discussed in Sec. II.B, Eq. (24) is defined and physically relevant over $1 \le \eta \le \eta_{\max}$. In this range, Fig. 1 indicates it roughly resembles a 'skew parabola' in $\eta$ (that becomes unbounded at $\eta = \eta_{\max}$) over a wide range of pressures. Consistent with this geometric interpretation, a useful form of $\varphi$ is given by

$$\varphi(\eta) = c_1 \left[ c_2 + \frac{(\eta - c_3)^2}{\eta_{\max} - \eta} \right], \tag{33}$$

where $c_1$-$c_3$ are parameters that may be determined by implementing some principal features of Eq. (24) as outlined in Sec. II.B:

1) $K_S$ at $\eta = 1$ is given by Eq. (26),

2) The $\eta$-derivative of $K_S$ (which appears to be nearly constant for a significant range of $\eta > 1$) at $\eta = 1$ is given by Eq. (27),

3) $K_S$ has a minimum between $\eta = 0$ and $\eta = \eta_{\max}$.

With Eqs. (26) and (27), constraints (1) and (2) result in the conditions

$$\frac{(4s-1)P}{2} + \rho_{\text{ref}} c_{\text{ref}}^2 = P c_1 \left[ c_2 + \frac{(1-c_3)^2}{\eta_{\max} - 1} \right], \tag{34}$$

$$\frac{4s-1}{4}\left(-P + 2\rho_{\text{ref}} c_{\text{ref}}^2\right) = P \frac{c_1 (c_3 - 1)(1 + c_3 - 2\eta_{\max})}{(\eta_{\max} - 1)^2}. \tag{35}$$

Assuming $P >> \rho_{\text{ref}} c_{\text{ref}}^2$ , Eqs. (34) and (35) become

$$\frac{(4s-1)}{2} = c_1 \left[ c_2 + \frac{(1-c_3)^2}{\eta_{\max} - 1} \right], \tag{36}$$

$$\frac{1-4s}{4}=\frac{c_1\left(c_3-1\right)\left(1+c_3-2\eta_{\max}\right)}{\left(\eta_{\max}-1\right)^2}. \tag{37}$$

Equations (36) and (37) will prove most accurate in strong shock scenarios where the shock pressure $P$ is much larger than an unshocked 'reference pressure' associated with $\rho_{\text{ref}}$ and $c_{\text{ref}}$, but not for cases where $P \to 0$ (which are not of principal interest to this work).

Constraint (3) provides the remaining relation to close Eqs. (36) and (37). With Eq. (33), it is readily observed that the parameter $c_3$ may be interpreted as the $\eta$-position of the minimum in $K_S$. As discussed in Sec. II.B, the location of this minimum varies with $P$, but falls between $\eta = 1$ and $\eta = \eta_{\max}$. As the $P$-dependence of $c_3$ cannot be captured using Eq. (33), the following approximation will be used throughout the remainder of this work:

$$c_3 = 1 + q\left(\eta_{\max}-1\right), \tag{38}$$

i.e., the minimum in $K_S$ is taken to occur at an arbitrary fractional position $q$ ($0 \le q \le 1$) within the applicable $\eta$-range. With Eqs. (25) and (36)-(38), the constants $c_1$-$c_3$ appearing in Eq. (33) are

$$c_1 = \frac{1-4s}{4q\left(q-2\right)\left(s-1\right)}, \tag{39}$$

$$c_2 = q\left[4\left(s-1\right)-q\left(2s-1\right)\right], \tag{40}$$

$$c_3 = \frac{q+s-1}{s-1}. \tag{41}$$

Equation (33) is then depicted in Fig. 2 for various values of $P$.

Figure 2 shows that Eqs. (30) and (33) are qualitatively similar to Eq. (24) for a wide range of pressures. Closer agreement for any given P is controlled by the parameter q; in Fig. 2 the choice $q = 0.25$ corresponds to close agreement when $P = 10\rho_{\text{ref}}c^2_{\text{ref}}$. In this sense the parameter $q$ is a 'knob' that may be selected consistent with the expected pressure history of a solution under investigation.

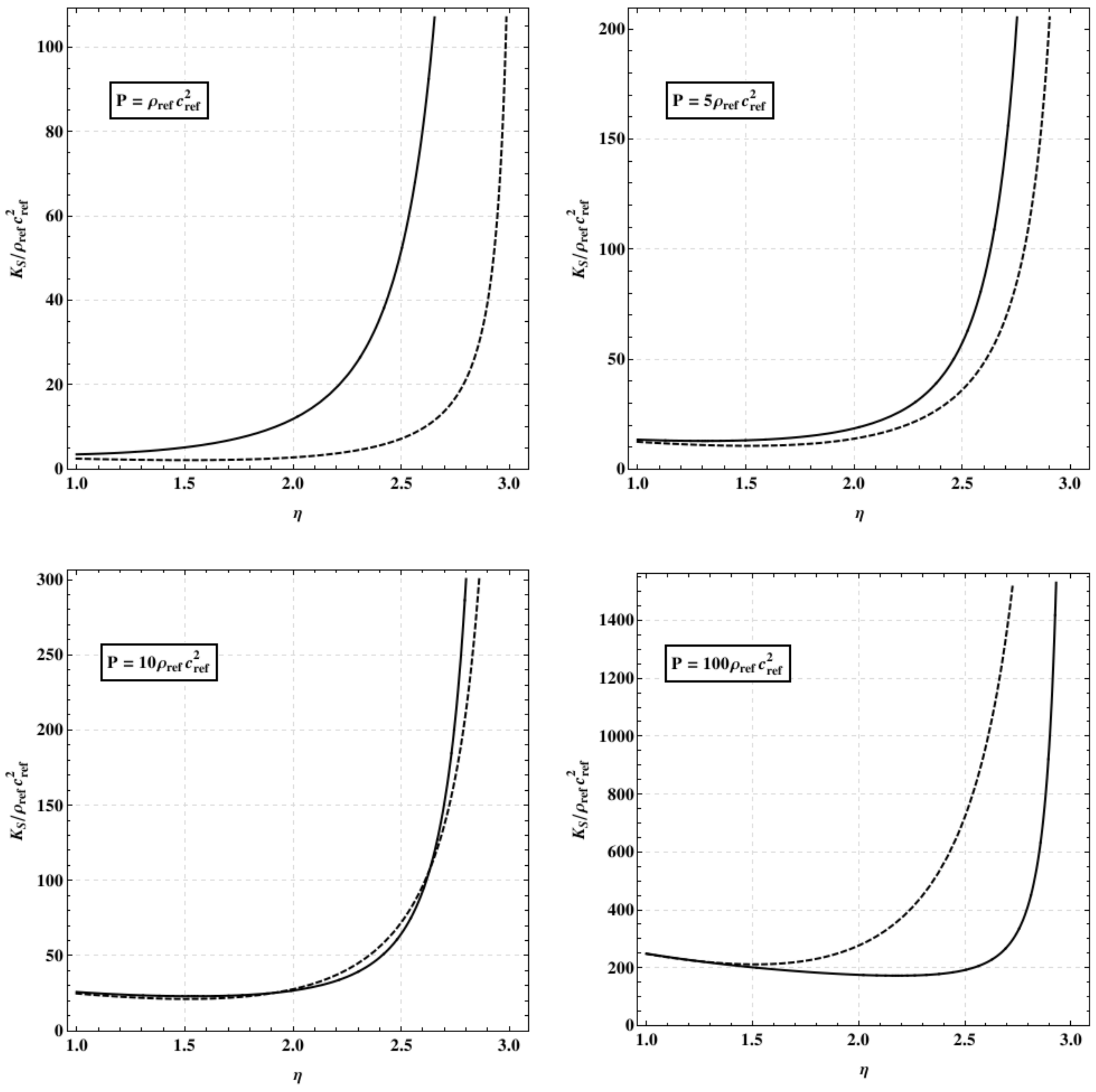

**Figure 2: Adiabatic bulk modulus for M-G EOS (solid line) and scale-invariant approximation (dashed line), *s* = 1.489, *q* = 0.25.**

### III.B. Equation of State

The scale-invariant approximate EOS corresponding to Eq. (39) is given by Eq. (31), with the Grüneisen parameter $\Gamma$ constructed via Eq. (32). This relation becomes

$$c_1\left[c_2+\frac{(\eta-c_3)^2}{\eta_{\max}-\eta}\right]=1+\Gamma(\eta)+\eta\frac{\Gamma'(\eta)}{\Gamma(\eta)}, \tag{42}$$

which has no known closed-form solution, but may be manipulated to show that $\Gamma \geq 0$ for $1 \leq \eta < \eta_{\max}$ (see Appendix A). It may also be readily integrated numerically subject to an initial condition obtained by requiring that Eqs. (22) and (31) are equivalent at some specified $\eta_i$,

$$P_{\mathrm{H}}(\eta_i)+2(2s-1)\rho_{\mathrm{ref}}\frac{\eta_i}{1+\eta_i}\left[e-e_{\mathrm{H}}(\eta_i)\right]=\rho_{\mathrm{ref}}\eta_i\Gamma(\eta_i)(e-e_{\mathrm{ref}}). \tag{43}$$

For Eq. (43) to be satisfied it must be an identity in powers of $e$, so that

$$\Gamma(\eta=\eta_i)=\frac{2(2s-1)}{1+\eta_i}, \tag{44}$$

and

$$\begin{aligned} e_{\mathrm{ref}} &= e_{\mathrm{H}}(\eta_i)-\frac{(1+\eta_i)P_{\mathrm{H}}(\eta_i)}{2(2s-1)\rho_{\mathrm{ref}}\eta_i} \\ &= \frac{(s-1)\eta_i-s}{(2s-1)\rho_{\mathrm{ref}}\eta_i}P_{\mathrm{H}}(\eta_i). \end{aligned} \tag{45}$$

Equation (44) is the required initial condition for the numerical integration of Eq. (42) for $\Gamma$; Eq. (45) must be simultaneously satisfied for the specified $\eta_i$, and thus sets the necessary value of the integration constant $e_{\mathrm{ref}}$. Since Eq. (45) indicates that $e_{\mathrm{ref}} \leq 0$ for $1 \leq \eta < \eta_{\max}$, Eq. (31) results in $P \geq 0$ since $\Gamma \geq 0$ for $1 \leq \eta < \eta_{\max}$.

While in principle either $\eta_i$ or $e_{\mathrm{ref}}$ may be chosen arbitrarily (but within the bounds of physical realism; i.e., $1 \leq \eta_i \leq \eta_{\max}$), in this work an initial condition at $\eta_i = 1$ will be used (so as to ensure a non-negative energy state in conjunction with $P = 0$ and $\rho > 0$). In this case, Eqs. (44) and (45) become, respectively,

$$\Gamma(\eta_i=1)=2s-1, \tag{46}$$

$$e_{\mathrm{ref}}=0, \tag{47}$$

so that Eq. (31) becomes

$$P = \rho_{\text{ref}} \eta \Gamma(\eta) e, \tag{48}$$

where $\Gamma$ is determined by a numerical solution of Eqs. (42) and (46).

## IV. CONVERGING SHOCK WAVE

The symmetry analysis of Eqs. (1), (2), and (7) that yields Eq. (30) may also be used to construct a system of similarity variables through which Eqs. (1), (2), and (7) may be reduced to ordinary differential equations (ODEs). An appropriate change to dimensionless variables is

$$\xi = \left( \frac{t}{t_{\text{ref}}} \right)^{-\alpha} \frac{r}{c_{\text{ref}} t_{\text{ref}}}, \tag{49}$$

$$D(\xi) = \frac{\rho}{\rho_{\text{ref}}}, \tag{50}$$

$$V(\xi) = \left( \frac{t}{t_{\text{ref}}} \right)^{1-\alpha} \frac{u}{c_{\text{ref}}}, \tag{51}$$

$$\Pi(\xi) = \left( \frac{t}{t_{\text{ref}}} \right)^{2(1-\alpha)} \frac{P}{\rho_{\text{ref}} c_{\text{ref}}^2}, \tag{52}$$

where $\alpha$ and $t_{\text{ref}}$ are constants to be determined. With Eqs. (49)-(52), Eq. (30) for the scale-invariant adiabatic bulk modulus likewise becomes

$$K_S = \rho_{\text{ref}} c_{\text{ref}}^2 \left( \frac{t}{t_{\text{ref}}} \right)^{2(\alpha-1)} \Pi \varphi(D), \tag{53}$$

and expressions for the various derivatives appearing in Eqs. (1), (2), and (7) are given by

$$\begin{aligned} \frac{\partial}{\partial r} &= \frac{\partial \xi}{\partial r} \frac{d}{d\xi} \\ &= \frac{1}{c_{\text{ref}} t_{\text{ref}} \left( \frac{t}{t_{\text{ref}}} \right)^{\alpha}} \frac{d}{d\xi}, \end{aligned} \tag{54}$$

$$\begin{aligned} \frac{\partial}{\partial t} &= \frac{\partial \xi}{\partial t} \frac{d}{d\xi} \\ &= -\alpha \xi \left( \frac{t}{t_{\text{ref}}} \right)^{-1} \frac{d}{d\xi}. \end{aligned} \tag{55}$$

Substituting Eqs. (49)-(55) into Eqs. (1), (2), and (7) then yields the ODEs

$$\left(V-\xi\right)D'+D\left(V'+\frac{kV}{\xi}\right)=0\,, \tag{56}$$

$$\left(V-\xi\right)V'-\left(\frac{1}{\alpha}-1\right)V+\frac{\Pi'}{D}=0\,, \tag{57}$$

$$\left(V-\xi\right)\Pi'-2\left(\frac{1}{\alpha}-1\right)\Pi+\Pi\varphi\left(D\right)\left(V'+\frac{kV}{\xi}\right)=0\,, \tag{58}$$

where the primes denote $\xi$-derivatives.

Equations (56)-(58) are valid for any scale-invariant flow captured by Eqs. (1), (2), and (7). For flows featuring shock waves, the Rankine-Hugoniot conditions must be similarly transformed. Substituting Eqs. (48)-(52) into the Rankine-Hugoniot conditions given by Eqs. (13)-(15) yields

$$-u_{\mathrm{s}}\rho_{\mathrm{ref}}=\left[c_{\mathrm{ref}}\left(\frac{t}{t_{\mathrm{ref}}}\right)^{\alpha-1}V_{\mathrm{s}}-u_{\mathrm{s}}\right]\rho_{\mathrm{ref}}D_{\mathrm{s}}\,, \tag{59}$$

$$\rho_{\mathrm{ref}}c_{\mathrm{ref}}^{2}\left(\frac{t}{t_{\mathrm{ref}}}\right)^{2(\alpha-1)}\Pi_{\mathrm{s}}-u_{\mathrm{s}}\rho_{\mathrm{ref}}c_{\mathrm{ref}}\left(\frac{t}{t_{\mathrm{ref}}}\right)^{\alpha-1}V_{\mathrm{s}}=0\,, \tag{60}$$

$$\begin{aligned}\frac{1}{2}u_{\mathrm{s}}^{2}&=\rho_{\mathrm{ref}}c_{\mathrm{ref}}^{2}\left(\frac{t}{t_{\mathrm{ref}}}\right)^{2(\alpha-1)}\frac{\Pi_{\mathrm{s}}}{\rho_{\mathrm{ref}}D_{\mathrm{s}}\Gamma\left(D_{\mathrm{s}}\right)}\\&+\rho_{\mathrm{ref}}c_{\mathrm{ref}}^{2}\left(\frac{t}{t_{\mathrm{ref}}}\right)^{2(\alpha-1)}\frac{\Pi_{\mathrm{s}}}{\rho_{\mathrm{ref}}D_{\mathrm{s}}}+\frac{1}{2}\left[c_{\mathrm{ref}}\left(\frac{t}{t_{\mathrm{ref}}}\right)^{\alpha-1}V_{\mathrm{s}}-u_{\mathrm{s}}\right]^{2},\end{aligned} \tag{61}$$

where the subscript 's' indicates a quantity is evaluated at $\xi_{\mathrm{s}}=\xi(r=r_{\mathrm{s}})$; Eqs. (59)-(61) must be satisfied for all time, indicating that the shock speed $u_{\mathrm{s}}$ must obey

$$u_{\mathrm{s}}=c_{\mathrm{ref}}\left(\frac{t}{t_{\mathrm{ref}}}\right)^{\alpha-1}. \tag{62}$$

Integrating Eq. (62) yields the scale-invariant shock position,

$$r_{\mathrm{s}}=c_{\mathrm{ref}}t_{\mathrm{ref}}\left(\frac{t}{t_{\mathrm{ref}}}\right)^{\alpha}, \tag{63}$$

where in Eqs. (62) and (63), proportionality and integration constants have been judiciously selected so as to yield the indicated forms. If $t_{\mathrm{ref}} < 0$, $t < 0$, and $0 < \alpha < 1$ in Eqs. (62) and (63), the shock wave accelerates toward the origin as $t \rightarrow 0$ from negative times.

Moreover, with Eqs. (49) and (63) it follows that $\xi_{\mathrm{s}} = 1$, and Eqs. (59)-(61) become

$$-1 = (V_s - 1) D_s , \tag{64}$$

$$\Pi_s - V_s = 0 , \tag{65}$$

$$\frac{1}{2} = \frac{\Pi_s \left[ 1 + \Gamma(D_s) \right]}{D_s \Gamma(D_s)} + \frac{1}{2}(V_s - 1)^2 , \tag{66}$$

which may be solved numerically for $D_s$, $V_s$, and $\Pi_s$, which in turn may be interpreted as initial conditions for the numerical solution of the first order ODEs given by Eqs. (56)-(58)*. In practice, Eqs. (56)-(58) are solved numerically from $\xi = 1$ (denoting the shocked fluid state immediately adjacent to the shock wave) to $\xi \to \infty$ (denoting the fluid state as $r \to \infty$ for finite times). For $\xi < 1$ (denoting the fluid state of the unshocked region into which the shock wave is propagating), Eqs. (62)-(64) trivially reduce to

$$D = 1 , \tag{67}$$

$$\Pi = 0 , \tag{68}$$

$$V = 0 , \tag{69}$$

consistent with the assumed unshocked state ($\rho = \rho_{\mathrm{ref}}$, $P = 0$, and $u = 0$ for $r < r_s$).

Equations (56)-(58) with the solution of Eqs. (64)-(66) as initial conditions are equivalent to the system discussed by Boyd et al.[22]; it may have a physical solution only in the case that $\alpha$ is regarded as a nonlinear eigenvalue, selected so that the solution of the system is never singular for $1 \leq \xi < \infty$. For a given geometry and EOS parameterization this requirement sets a unique value for the parameter $\alpha$, as demonstrated in the context of the ideal gas by Guderley[3], Butler[49], Lazarus[14], Ramsey et al.[50], and many other authors. Consistent with prior work, a solution of the eigenvalue problem for $\alpha$ will be constructed in the context where $\varphi$ appearing in Eqs. (56)-(58) is given by Eq. (33).

## IV.A. Numerical Example

With Eq. (33), Eqs. (56)-(58) become, respectively,

$$(V - \xi) D' + D\left( V' + \frac{kV}{\xi} \right) = 0 , \tag{70}$$

$$(V - \xi) V' - \left( \frac{1}{\alpha} - 1 \right) V + \frac{\Pi'}{D} = 0 , \tag{71}$$

$$(V - \xi)\Pi' - 2\left( \frac{1}{\alpha} - 1 \right)\Pi + c_1 \left[ c_2 + \frac{(D - c_3)^2}{\eta_{\max} - D} \right] \Pi \left( V' + \frac{kV}{\xi} \right) = 0. \tag{72}$$

*The question of whether or not solutions exist for this ODE system for arbitrary $\varphi(D)$ has yet to be definitely answered. It is known that solutions exist for $\varphi$ = const. > 1; another solution will be demonstrated to exist in Sec. IV.A.

For Eqs. (70)-(72) to be solved numerically subject to the initial conditions given by the solution of Eqs. (64)-(66), the approximate EOS constant $s$ and $q$ must be specified (as must the dimensional constants $\rho_{ref}$ and $c_{ref}$, but only for the transformation of the solution of Eqs. (70)-(72) back to physical variables). As an example, copper (Cu) is well-characterized by a M-G EOS, and its parameterization is given by Cooper[44] as:

$\rho_{ref}$ = 8.930 g/cm$^3$,
$c_{ref}$ = 3.940 km/s,
$s$ = 1.489.

Moreover, the parameter $q$ will be taken to assume the value

$q$ = 0.25.

With this parameterization, Eqs. (70)-(72) subject to the solution of Eqs. (64)-(66) as initial conditions may be solved by a variety of numerical integration packages for ODEs. Using Wolfram Mathematica 9[51], it is found via a shooting method that

$\alpha$ = 0.517778,

enables the existence of a non-singular solution. An example of the corresponding solution transformed back to physical variables [using Eqs. (49)-(52)] is depicted in Figs. 3 and 4. The flow field depicted in Fig. 4 echoes many of the features of the ideal gas Guderley solution (see, for example, Zel'dovich and Raizer[4] or Ramsey et al.[50]), including a pressure field with a maximum, and a density field that grows with $r$. Indeed, as $r \rightarrow \infty$, the solution achieves a maximum density of approximately $\rho_{max}$ = 25.4 g/cm$^3$, which is less than $\rho_{ref}\,\eta_{max}$ = 27.2 g/cm$^3$ as given by Eq. (27) for the M-G EOS.

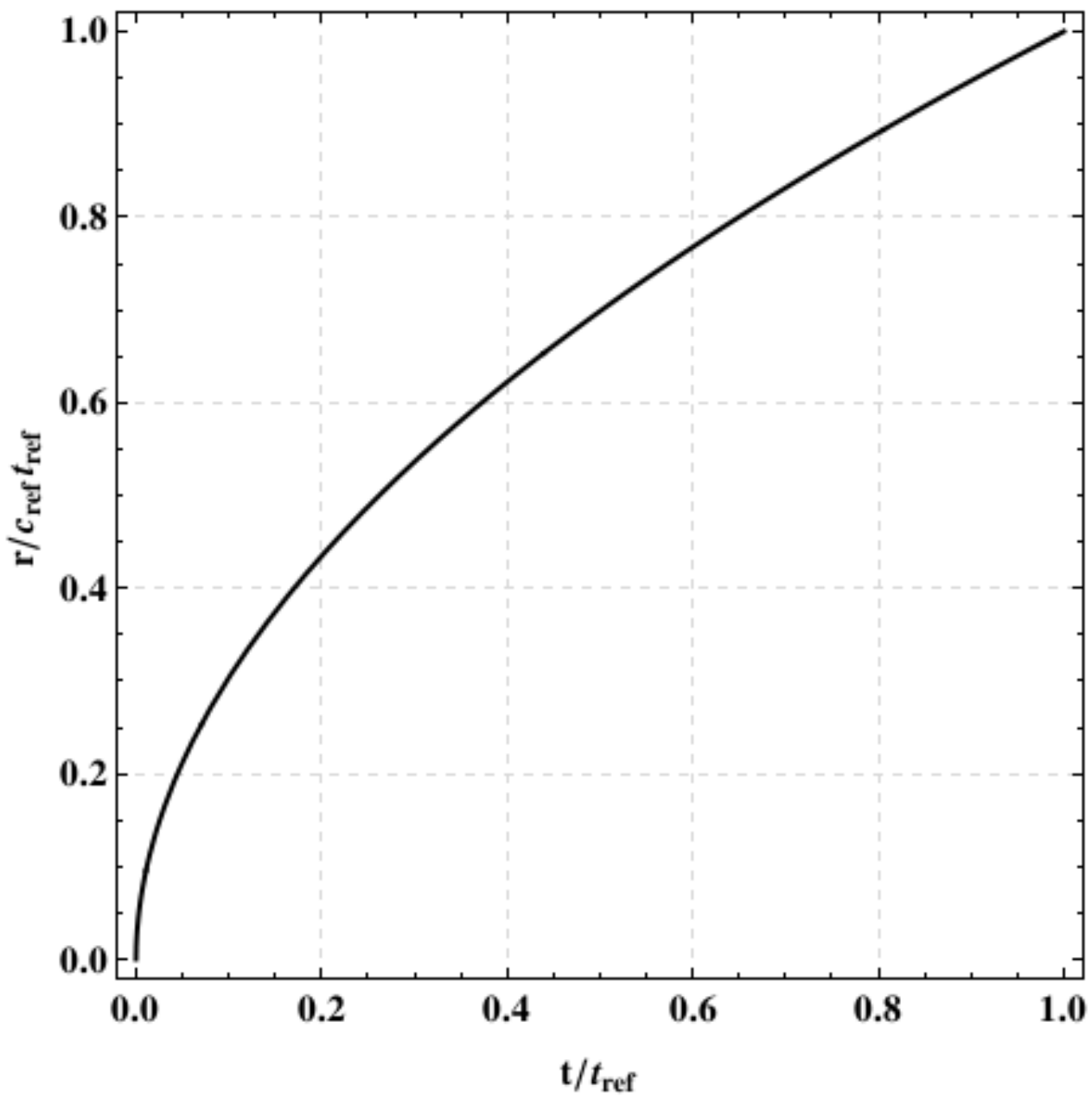

**Figure 3: Converging shock trajectory for Guderley-like problem with approximate M-G EOS.**

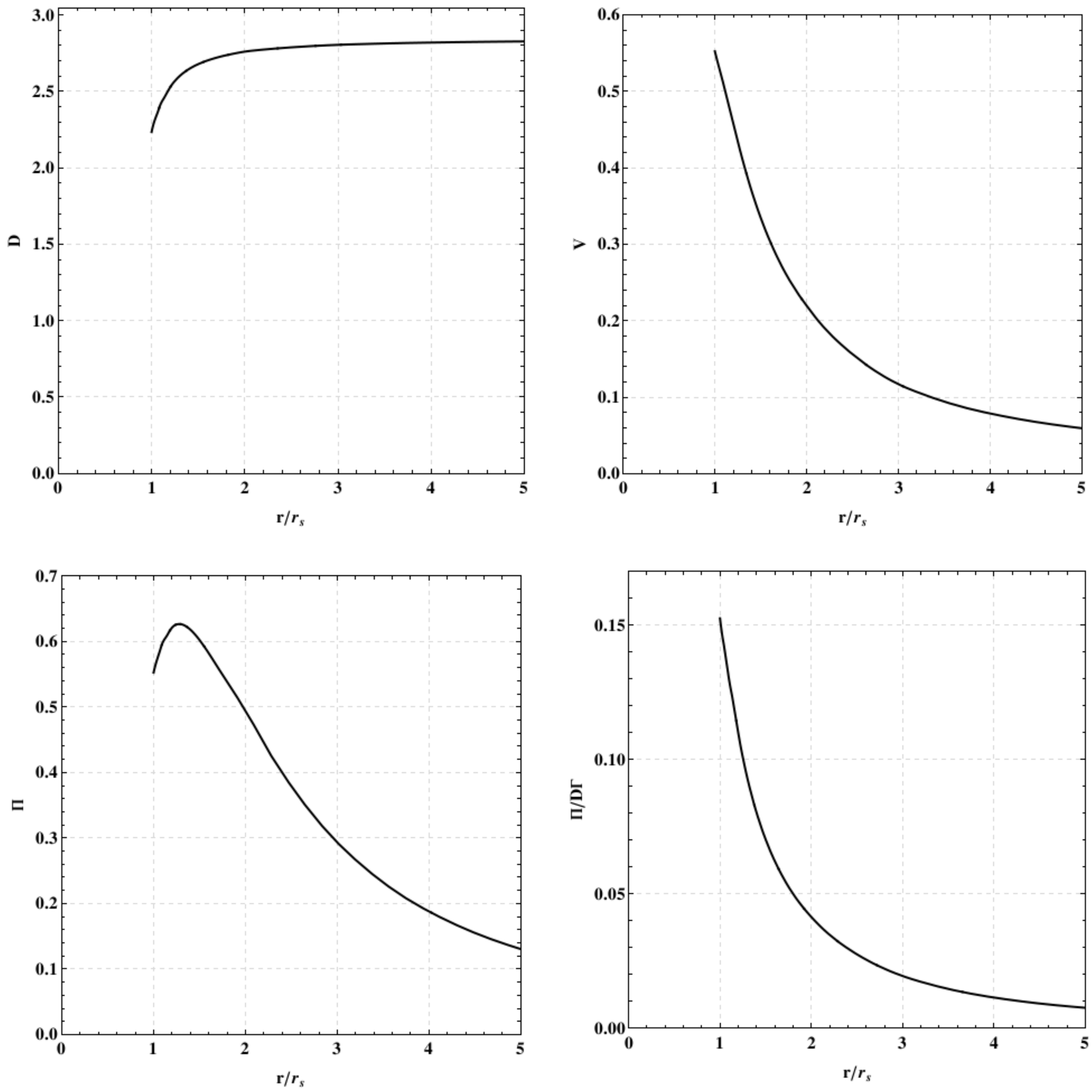

**Figure 4: Converging flow field for Guderley-like problem with approximate M-G EOS. Clockwise from top left: density, velocity, SIE, pressure.**

To further assess the validity of the solution results, Fig. 5 depicts as a function of space and time the relative difference between the M-G [Eq. (24)] and approximate [Eqs. (30) and (33)] adiabatic bulk moduli, calculated using the solutions for $\rho$ and $P$ found via Eqs. (70)-(72) [and transformed back to physical variables using Eqs. (49)-(52)]. Figure 5 thus allows for the identification of space-time domains where the solution computed using the scale-invariant bulk modulus falls within a given accuracy relative to the actual M-G EOS. As may be expected in light of Fig. 2, the approximate solution is most accurate only within a restricted space-time domain.

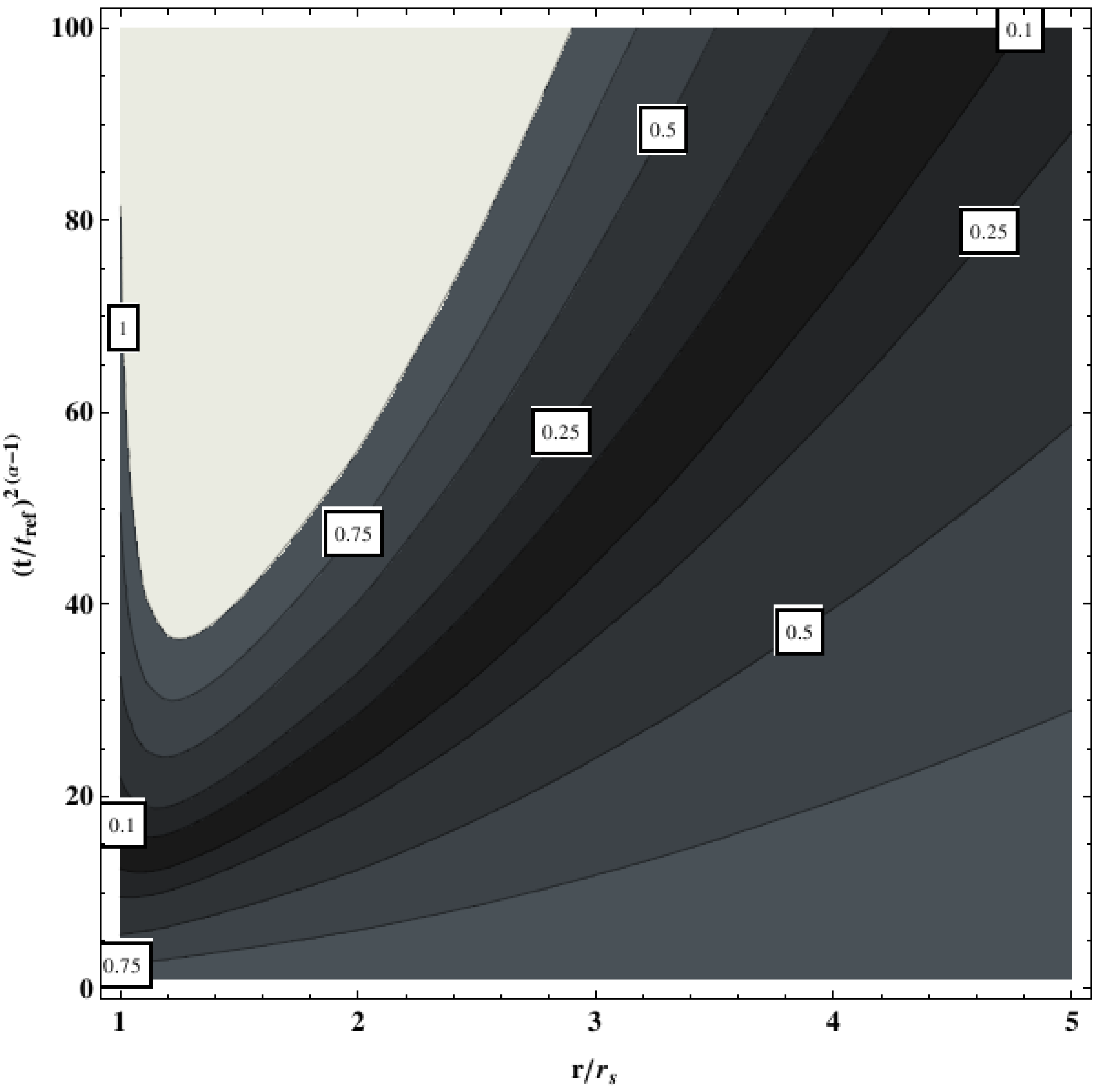

**Fig. 5: Relative difference between Eqs. (24) and (30)/(33), with $\rho$ and $P$ calculated as functions of $r$ and $t$ via Eqs. (70)-(72). Position is increasing from $r = r_s$ along the horizontal axis, and time is decreasing from $t = t_{ref}$ to $t \to 0$ along the vertical axis. Contours denote curves of constant relative difference.**

## V. CONCLUSIONS

This work includes a brief review of some properties of a M-G EOS that is expected to be applicable to the characterization of shock compressed crystalline solids. However, this EOS is not of a form compatible with the construction of scale-invariant similarity solutions of the compressible Euler equations, including Guderley-like converging shock flows. This incompatibility naturally arises from the presence of multiple inherent dimensional scales within the M-G EOS; as such, hydrodynamic scaling phenomena should not be expected to exist in this context.

Using an outcome of symmetry analysis as applied to the compressible Euler equations, scale-invariant approximations to the M-G EOS can be constructed, and then used to determine Guderley-like or other solutions. In this work, notable geometric features of the adiabatic bulk modulus corresponding to the M-G EOS are used motivate the approximate scale-invariant form. The resulting numerical solution for a converging shock flow is then readily computed, and then passed back through the exact and approximate bulk moduli to quantitatively assess the quality of the approximation. This procedure also allows for the determination of regions within the approximate solution field that fall within a prescribed accuracy.

Moreover, because of its relevance to the M-G EOS and specific classes of physical flows, the numerical Guderley-like solution computed in Sec. IV.A may also prove useful as a test problem for the quantitative verification of compressible Euler codes. The majority of such studies have historically been performed in the context of the ideal gas EOS (see, for example, Ramsey, et al.[50]), and an extension to even a fictitious (but physically motivated) non-ideal material such as that presented in this work opens the door to the construction of increasingly discriminating and relevant test problems.

The preceding work may be extended in a variety of ways:

- Particular forms of the reference pressure/energy and Grüneisen parameter appearing in the M-G EOS are used throughout this work. As discussed in Sec. II.B, many additional parameterizations exist and could be similarly investigated.

- A simple functional form of the approximate adiabatic bulk modulus [Eqs. (30) and (33)] was selected using geometric arguments arising from inspection of Fig. 1. Increasingly elaborate scale-invariant forms could be investigated.

- Additional scale-invariant flows aside from the Guderley converging shock (e.g., the Noh[46] or Sedov[1,2] problems) could be investigated in conjunction with the approximate EOS methodology.

- Quasi-similar methods (as discussed in Sec. I) may be employed to more rigorously quantify the departure of non-similarity flows from their self-similar limits.

## ACKNOWLEDGEMENTS

This work was performed under the auspices of the United States Department of Energy by Los Alamos National Security, LLC, at Los Alamos National Laboratory under contract DE-AC52-06NA25396. The authors thank E.J. Albright, D. Garcia, M. Haertling, L. Margolin, J. McHardy, R. Pelak, and J. Schmidt for valuable insights on these topics.

## A. GRÜNEISEN PARAMETER FOR THE SCALE-INVARIANT EQUATION OF STATE

For a scale-invariant adiabatic bulk modulus of the form given by Eq. (30), the Grüneisen parameter $\Gamma$ appearing in the corresponding EOS is given by a solution of Eq. (32). While it is not expected that Eq. (32) will yield a closed-form result for almost any choice of $\varphi$, this relation may be manipulated to reveal some key features of $\Gamma$.

For arbitrary $\varphi$, the solution of Eq. (32) may be expressed in terms of quadratures as

$$\Gamma(\eta) = \frac{\iota'(\eta)}{C_1 + \iota(\eta)}, \tag{A.1}$$

where

$$\iota(\eta) = \int_1^{\eta} \frac{1}{x_1} \exp\left( \int_1^{x_1} \frac{\varphi(x_2) - 1}{x_2} dx_2 \right) dx_1, \tag{A.2}$$

$C_1$ is an arbitrary constant, and primes denote derivatives with respect to the indicated argument. Equation (A.2) may be equivalently re-expressed as

$$\eta \frac{d}{d\eta} \ln\left[\eta \iota'(\eta)\right] = \varphi(\eta) - 1, \tag{A.3}$$

or, with Eq. (33),

$$\eta \frac{d}{d\eta} \ln\left[\eta \iota'(\eta)\right] = c_1 \left[ c_2 + \frac{(\eta - c_3)^2}{\eta_{\max} - \eta} \right] - 1. \tag{A.4}$$

The first integral of Eq. (A.4) is then given by

$$\iota'(\eta) = C_2 \eta^{c_4} \left(\eta_{\max} - \eta\right)^{c_5} \exp(-c_1 \eta), \tag{A.5}$$

where

$$c_4 = \frac{c_1 c_3^2}{\eta_{\max}} + c_1 c_2 - 1, \tag{A.6}$$

$$c_5 = \frac{c_1 \eta_{\max}^2 - 2 c_1 c_3 \eta_{\max} + c_1 c_3^2}{\eta_{\max}}, \tag{A.7}$$

and $C_2$ is an arbitrary integration constant.

While Eq. (A.5) still has no known closed-form solution, with Eq. (A.1) it provides useful information on Γ. In particular, Eq. (A.1) reveals that $\iota(\eta) \geq 0$ for $1 \leq \eta < \eta_{max}$. Moreover, for $C_2 > 0$, Eq. (A.5) reveals that $\iota'(\eta) > 0$ for $1 \leq \eta < \eta_{max}$. Thus, for the appropriate choice of the integration constants $C_1$ and $C_2$, Eq. (A.1) yields $\Gamma \geq 0$ for $1 \leq \eta < \eta_{max}$.